\documentclass{article}
\usepackage[margin=1in]{geometry}
\usepackage{graphicx}
\usepackage{float}

\title{Narrative Review of Emotional Expression Support in XR: Psychophysiology of Speech-to-Text Interfaces}

\author{Sunday D. Ubur, Denis Gracanin}

\date{}

\begin{document}

\maketitle

\begin{abstract}
This narrative review examines recent advancements, limitations, and research gaps in integrating emotional expression into speech-to-text (STT) interfaces within extended reality (XR) environments. Drawing from 37 peer-reviewed studies published between 2020 and 2024, we synthesized literature across multiple domains—including affective computing, psychophysiology, captioning innovation, and immersive human-computer interaction. Thematic categories include communication enhancement technologies for Deaf and Hard of Hearing (DHH) users, emotive captioning strategies, visual and affective augmentation in AR/VR, speech emotion recognition, and the development of empathic systems. Despite the growing accessibility of real-time STT tools, such systems largely fail to convey affective nuance, limiting the richness of communication for DHH users and other caption consumers. This review highlights emerging approaches such as animated captions, emojilization, color-coded overlays, and avatar-based emotion visualization, but finds a persistent gap in real-time emotion-aware captioning within immersive XR contexts. We identify key research opportunities at the intersection of accessibility, XR, and emotional expression, and propose future directions for the development of affect-responsive, user-centered captioning interfaces.
\end{abstract}

\vspace*{.1in}
\noindent\textbf{Keywords:} Emotional expression, Psychophysiology, Speech-to-text, Empathic machine, Emotion-aware interfaces, Accessible communication, AR/VR captioning, Multimodal interaction, Human-computer empathy, Neurodiversity, DHH technology.

\section{Introduction}
In the ever-evolving landscape of communication technologies, Speech-to-Text (STT) interfaces play a pivotal role in enhancing accessibility and inclusivity, particularly for individuals with hearing impairments. The ability to convert spoken language into text facilitates communication in various contexts, ranging from online meetings to educational settings. However, amidst the strides made in improving accessibility, a critical aspect often overlooked is the incorporation of emotional expression into transcribed text generated by STT systems. Additionally, the review introduces a new thematic synthesis on visual and affective augmentation in captioning, highlighting emerging methods that combine real-time emotional recognition with user-centered visual design in immersive environments.

This study embarks on a comprehensive exploration of the existing literature surrounding emotional expression in STT interfaces, with a specific emphasis on the psychophysiology aspect within Extended Reality (XR). The objective is to identify advancements, limitations, and research gaps in the incorporation of emotional expression in transcribed text generated by STT systems. As communication technologies continue to advance, the ability to convey not only the semantic content but also the emotional nuances of speech becomes paramount for fostering richer and more meaningful interactions.

The ubiquity of STT applications, exemplified by tools like Live Transcribe, has significantly contributed to breaking communication barriers for the Deaf and Hard of Hearing (DHH) community. These applications offer real-time transcriptions that are invaluable in various scenarios, from professional communication to educational contexts. However, the inherent challenge lies in the loss of emotional nuance during the transcription process, posing a communication hurdle that this study seeks to address.

To unravel the complexities surrounding emotional expression in STT interfaces, our investigation spans various dimensions. From examining innovations in live transcription and closed captioning to delving into advancements in augmented reality (AR), emotive captioning, emotion recognition, and empathic machines, this study aims to provide a holistic understanding of the evolving tools and techniques. Additionally, the exploration extends into the realms of VR and AR, where unique opportunities arise to create immersive and emotionally resonant experiences, especially in the context of education and training.
In the subsequent sections, we detail the methodology employed in conducting the literature review, present the background encompassing communication enhancement technologies, and delve into innovations in captioning, emotion recognition, and empathic systems. The synthesis of these findings not only identifies current advancements but also points towards avenues for future research, emphasizing the ongoing need for inclusive and accessible communication technologies.

This paper contributes: (1) A narrative synthesis of 37 studies from 2020–2024; (2) A categorization of emotion integration methods in STT; (3) A synthesized review of visual and affective augmentation strategies in immersive captioning; (4) Identification of key research gaps in VR/AR-based emotional captioning; (5) Proposed research questions for future empirical investigation.

\section{Methods}

\paragraph{Search Strategy:} 
The search involved a comprehensive search of scholarly databases, including ACM Digital Library, IEEE Explore, Scopus, and Google Scholar, to identify relevant articles published between 2020 and 2024. The search terms used included ("Emotional Expressions" AND "Extended Reality" AND "Psychophysiology" AND "speech-to-text Interfaces" OR "Captions"), and the inclusion criteria encompassed studies related to emotional expressions in psychophysiology of STT interfaces. We searched individual databases, however, much of the data came from the Publish or Perish software \cite{harzing2007} as it extracts results across these databases mentioned.

\paragraph{Inclusion and Exclusion Criteria:}
The rest of the methods process was conducted using Rayyan software~\cite{ouzzani2016}.
Most of the papers included are peer-reviewed articles, books, and conference proceedings that provided insights into the background and direction of implementing emotional expressions in STT. Studies with a focus on generating emotional expressions in text were prioritized. Non-English language publications and articles without full-text availability were excluded.

\paragraph{Study Selection:}
The initial search yielded 1046 articles. After removing duplicates and conducting title and abstract screening, 37 articles were considered for full-text review. The final selection was based on the relevance of the content to the research question and the quality of the study design.

\paragraph{Data Extraction:}
Data extraction was performed to gather key information from the selected literature, including future work recommendations. This process involved summarizing findings, identifying common themes, and noting variations across studies.

\paragraph{Quality Assessment:}
Given the narrative nature of this review, a formal quality assessment was not conducted. However, efforts were made to critically appraised each study's methodology, sample size, and key findings to gauge the overall reliability of the evidence.

\paragraph{Synthesis:}
The synthesis involved organizing the selected literature into thematic categories and identifying overarching trends and patterns, and employed a narrative approach to present the key findings and critically discuss the implications of the literature on the research question.

\section{Background}

\begin{figure}[H]
    \centering
    \includegraphics[width=0.75\textwidth]{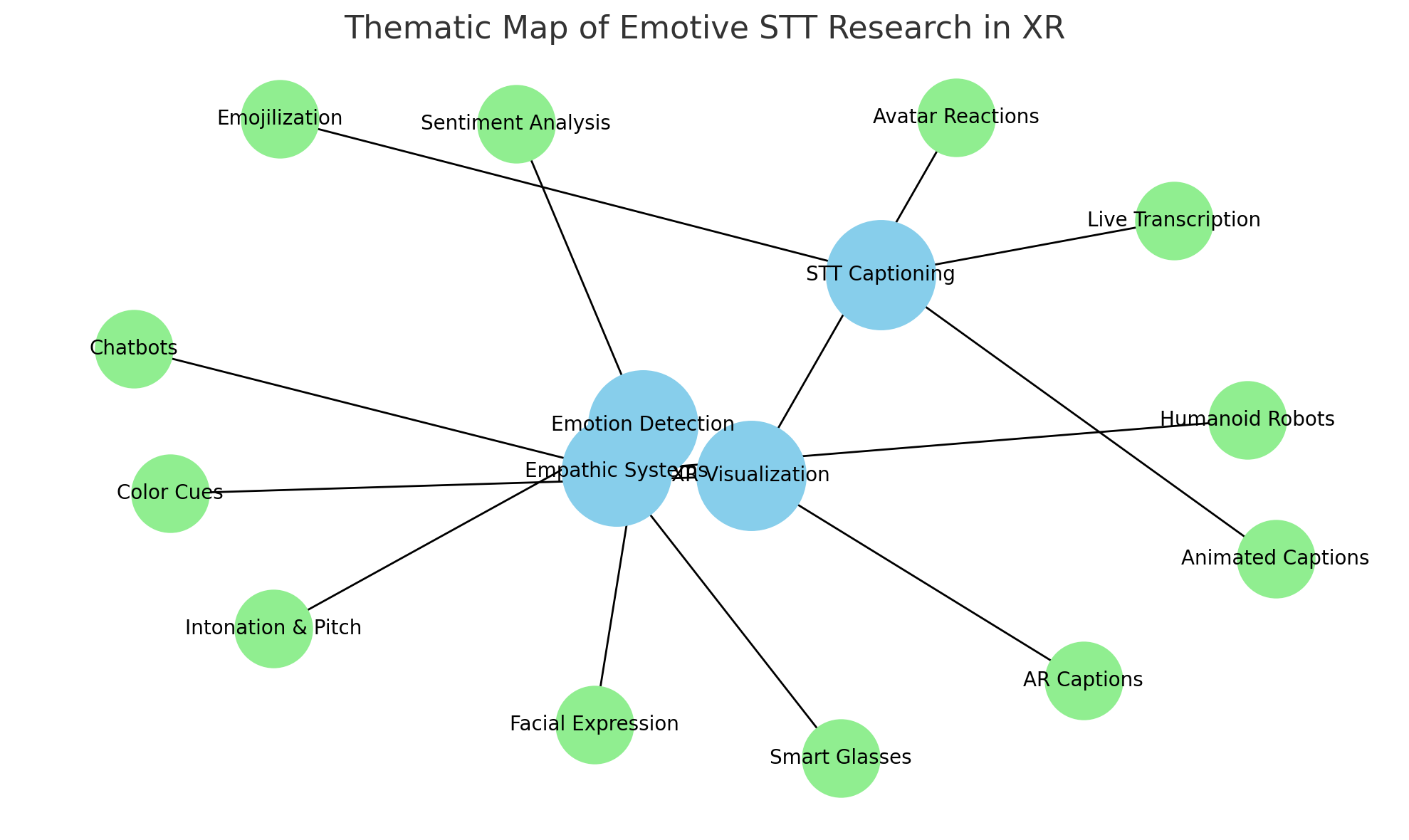}
    \caption{Thematic map summarizing the key areas reviewed in this study, including communication enhancement technologies, innovations in captioning, emotion recognition in XR, empathic machine interfaces, and emerging strategies in visual and affective augmentation.}
    \label{fig:thematic-map}
\end{figure}

This section provides the narrative review of the literature, with study limitations and comparisions among related studies.
\subsection{Communication Enhancement Technologies for DHH Individuals}
The exploration of communication enhancement technologies for DHH individuals has seen significant strides, particularly with the non-traditional applications of Live Transcribe, an automatic speech recognition (ASR) application \cite{loizides2020breaking}. This tool proves invaluable in various scenarios, from technical support and communication with colleagues to note-taking during meetings, especially during the prevalence of online meetings amid COVID-19 lockdowns. The adaptability of Live Transcribe for immobile DHH individuals by mounting the tablet on a boom mic stand is a notable advancement \cite{arnold1979interaction}, emphasizing its pivotal role in promoting inclusivity and accessibility for this demographic.

Closed captioning is a technology that has seen minimal innovation since its inception in the 1970s, however recent studies are shedding light on potential enhancements. One such is the animated text in captions \cite{rashid2008} which offers a promising avenue, providing improved access to emotive information often overlooked in traditional captions, such as music, sound effects, and intonation. This innovation, preferred by both hard of hearing and hearing participants, emphasizes the need to bridge the gap in conveying nonverbal nuances. This study and other related works was inspired by \cite{murphy1983impact} which delved into the impact of captions on the affective reactions of hearing-impaired children to television programming, revealing the potential of captions in enhancing emotional perception.

\subsection{Innovations in Captioning for Enhanced Emotional Communication}
Shifting focus to innovations in captioning for enhanced emotional communication, Rashid's framework \cite{rashid2006expressing} introduces the concept of using animation and standard properties to express basic emotions. This framework, associating emotions with animation properties, establishes a consistent method for applying animation to text captions. Similarly, to enhance emotions in captions \cite{lee2007emotive} incorporating graphics, color, and animation to illustrate sound and emotive information in television and film, a user study compared viewer reactions to video samples with emotive captions against conventional captioning, showcasing positive responses, particularly among hard of hearing viewers.

In the realm of Voice User Interfaces (VUI), Hu's paper \cite{hu2019emojilization} highlights the lack of emotional information in STT by proposing an emojilization tool. This tool automatically attaches emojis to generated text, compensating for emotional loss in the conversion process. The pilot study indicates that emojilized text enhances perceived emotions compared to plain text. 

We can also understand how to represent emotions in captions by analyzing how emotions are expressed in text. A study along this direction is the representation of emotions in text-based messaging applications \cite{poguntke2019smile} through visualizations in facial expression recognition in WhatsApp Web, and findings underscore users' preference for maintaining control over their emotions in private settings, emphasizing the need for exclusive presentation and sharing of emotions in certain contexts. An important application of emotional expressions in text and imagery on social media during the COVID-19 pandemic \cite{li2022text} adds a temporal dimension to the narrative, shedding light on how emotions are expressed in different modalities during challenging times.

\subsection{Emotion Recognition in AR and VR}

VR has gained attention in various fields due to its unique advantages in manipulating perceived scenarios and providing controlled experiences~\cite{kilteni2012sense}.
It has been found that VR can generate emotional reactions and experiences in users, as demonstrated by studies on 360 degree videos and VR educational tools~\cite{lie2023developing}.
In fact, emotional responses in VR games have been observed to be more intense compared to desktop games, both psychologically and physiologically~\cite{pallavicini2020virtual}.
This suggests that VR has the potential to enhance positive emotions, and it is worth to study if similar outcome can be experienced from STT in an AR or VR environment.

More studies have discussed advancements in AR technologies for DHH individuals, particularly in the classroom setting~\cite{alnafjan2017designing}.
The AR visual-captions system, implemented using Unity, ARKit, and AR Foundation, aims to provide real-time STT and visual elements around the teacher, creating an immersive learning experience.
The study not only introduced the prototype but also outlined future research directions, especially on  refining the design, conducting user studies, and extending the system's capabilities for group conversations.
Similarly, Real-time AR visual-captions~\cite{li2023real} for DHH children in classrooms builds on the AR-based system, providing a real-time solution for STT and keyword extraction, addressing the challenges faced by DHH children in mainstream education.

In higher education application, Pirker's exploration  into the potential of VR for computer science education~\cite{pirker2021potential} sheds light on the positive impact of VR environments on learning outcomes.
The user study reveals higher engagement, immersion, and positive emotions in the VR application compared to a web-based alternative, emphasizing VR's potential to enhance computer science education.
Given the importance of AR and VR in education training, making the virtual environment convey emotional expression is essential.
Chen's paper~\cite{chen2023emotion} focuses on facial expression recognition within immersive environments.
The proposed solution involves collecting real facial expression data using an infrared camera and light source, showcasing promising results for understanding human emotions in VR contexts.

\subsubsection{Visual and Affective Augmentation in XR}

\begin{figure}[ht]
    \centering
    \includegraphics[width=0.5\textwidth]{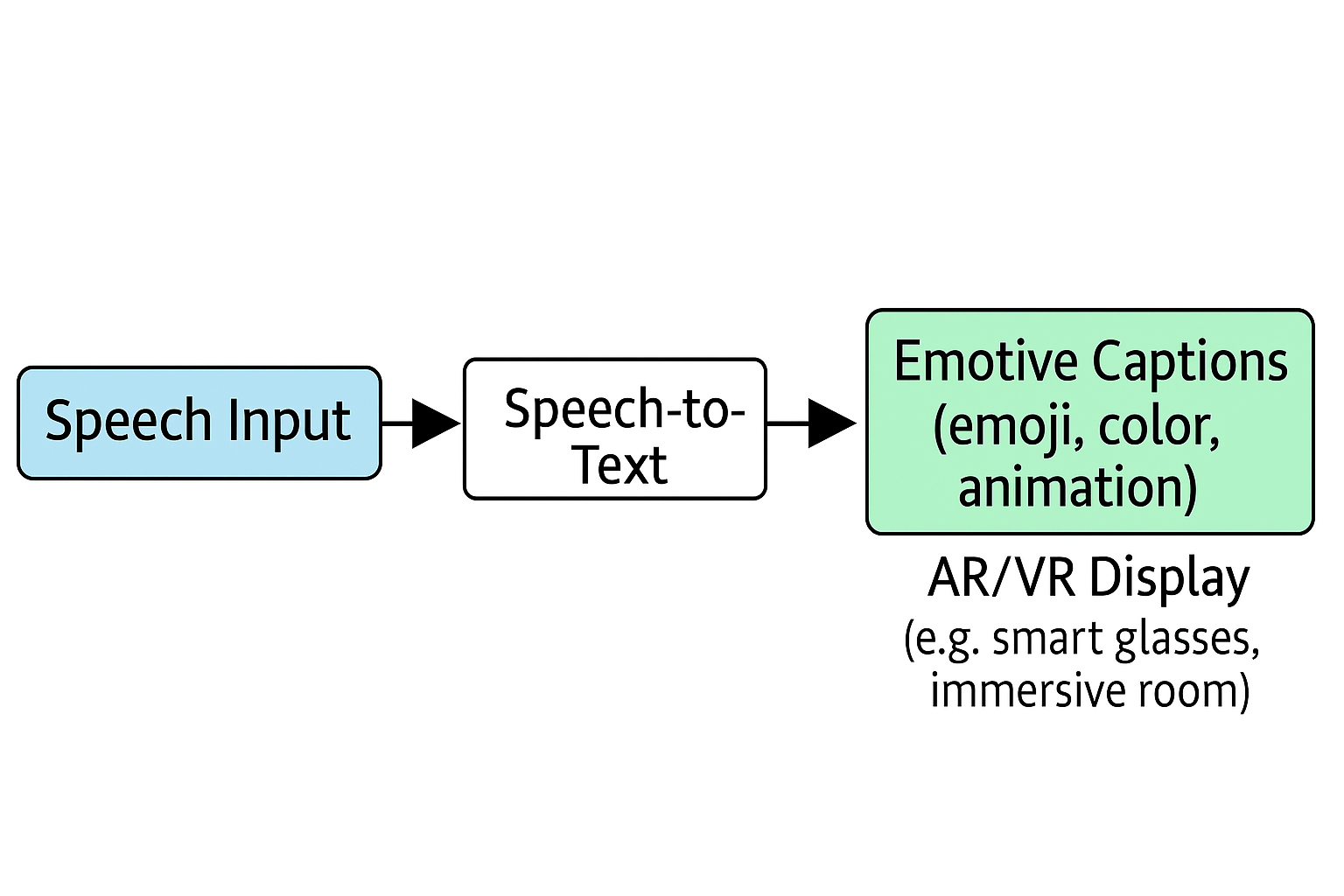}
    \caption{Workflow of an emotion-aware STT system in XR environments. Speech is converted to text, enhanced with emotional cues (e.g., emojis, color, animation), and displayed in AR/VR settings.}
    \label{fig:caption-workflow}
\end{figure}

As immersive AR and VR environments continue to gain traction in accessibility research, recent studies have emphasized the importance of incorporating visual and affective augmentations into captioning systems. These augmentations not only support emotional expression but also reduce cognitive strain and improve comprehension, particularly for DHH users and others who rely on captions in complex, real-time scenarios. A conceptual workflow of this integration is illustrated in Figure~\ref{fig:caption-workflow}, showing how STT outputs can be enhanced with emotional cues before delivery in XR displays.

Recent research highlights the cognitive demands of prolonged exposure to captions within VR HMDs, including increased eye strain and stress when captions are not personalized or well-positioned~\cite{10.1007/978-3-031-60881-0_24}. Participants expressed preferences for adjustable caption features such as text size, avatar use, and caption placement, underscoring the need for personalization in immersive captioning interfaces.

To support the design of emotionally expressive captions, a framework and accompanying interactive tool were developed to help designers visualize emotions in AR using abstract or representational visual elements~\cite{10.1145/3532525.3532527}. The framework guides decisions related to shape, animation, and spatial configuration of affective visuals, enabling real-time augmentation of emotional content in communication.

A related framework for AR-based affect visualization~\cite{10.1007/978-3-030-90176-9_49} builds on the dimensional model of emotion and focuses on ambiguous visual representations such as gradients, forms, and motion patterns. The study emphasizes that ambiguous, user-interpretable visuals can enhance emotional communication and user engagement without overwhelming or over-simplifying complex emotional states.

In shared VR environments, bi-directional emotion-sharing mechanisms have been shown to improve interpersonal trust, decision-making, and collaborative performance~\cite{10316483}. Visualizing emotions through simplified indicators such as valence icons contributed to more frequent emotional consensus and helped build rapport between unfamiliar collaborators, pointing to the potential of similar strategies in educational and captioning systems.

Color also plays a central role in emotional communication. One study explored mapping vocal emotional states to colored message bubbles~\cite{10.1145/3411763.3451698}, showing that users perceived emotional intensity more clearly through contextual color cues. However, negative emotional states sometimes reduced user comfort, highlighting the importance of careful calibration when applying visual augmentation to speech-based captions.

These findings are directly relevant to the development of accessible XR for education. A user-centric investigation into educational captioning for DHH individuals recommends embedding emotional context into captions through visual highlights, emojis, and keyword emphasis~\cite{10536263}. Such enhancements are not only beneficial for comprehension and engagement but also critical for reducing the sense of isolation and cognitive load associated with conventional captioning.

Collectively, these studies demonstrate that visual and affective augmentation is a necessary evolution in XR-based captioning. When combined with emotion recognition technologies and user-centered design, these strategies can transform how captions communicate meaning—enhancing not only what is said, but how it is felt.

\subsection{Speech Emotion Recognition: Techniques and Challenges}
An important technique found in the literature is the use of color coding and visualization of emotion in speech, which offer a unique perspective on conveying emotional information.  \cite{elor2021understanding} utilizes haptic feedback vests and immersive VR to quantify human emotion, providing insights for the development of emotionally relevant input devices. The use of bubble coloring in chat platforms \cite{bartram2017affective, cernea2015emotion} explores the visual representation of speech emotion, considering color alterations and participant feedback for conveying specific emotions effectively.
Other techniques in machine learning are being employed to include emotional recognition in text. One is Human-robot interaction and perception of emotions \cite{wang2004communicating}, demonstrating the effectiveness of animated text associated with emotional information in enhancing user interactions. The system's potential applications in healthcare and education underscore its versatile impact.

Further, Schiano's paper \cite{schiano2000face} delves into the perception of facial affect, bridging the gap between human facial expressions and prototype robot faces. The study informs the design of affective robot faces, contributing valuable insights into how emotions can be effectively communicated through facial expressions.In a related study,the analysis of humanoid avatar representations and emotions \cite{kumarapeli2022emotional} delves into the uncanniness factor, highlighting the importance of selecting appropriate avatar types for accurate expression communication. The study's findings support the need for avatar representations that effectively convey emotional cues in interactive systems. Finally, gestures can also be generated directly from speech using GANs \cite{rebol2021passing} thereby opening up new possibilities in human-computer interaction. The user study using Turing test-inspired evaluation demonstrates the potential of the proposed technique in creating realistic gestures from speech.
In conclusion, the narrative literature review traverses a spectrum of communication enhancement technologies, showcasing the evolution of tools and techniques to address the unique needs of DHH individuals. From innovations in live transcription and closed captioning to advancements in AR, emotive captioning, and speech emotion recognition, the literature reflects a dynamic landscape of research and development. The exploration of humanoid avatars, voice user interfaces, VR, human-robot interaction, and the color coding of speech emotions further emphasizes the interdisciplinary nature of this field. The review also underscores the significance of understanding emotional expressions in text, images, and gestures, offering a holistic perspective on communication technologies that cater to diverse sensory and cognitive needs. As we navigate the intricacies of enhancing communication for individuals with hearing impairments, these technological advancements pave the way for a more inclusive and accessible future.

\subsection{Empathic Machine}

Emotions serve as an implicit communication channel among humans, conveyed through spoken words, facial expressions, behavior, and physiological responses. This empathic form of communication enables individuals to recognize cues and respond empathetically. Despite computing devices excelling in understanding user context, there are ongoing challenges in the scientific community to create emotion-sensing and empathic applications for human-computer interaction \cite{bosch2022empathic}. In this section we look at the literature contributions in tackling these challenges and support empathic ability in machines.

\paragraph{Enhancing Empathic Abilities with Chatbots:}
The evolution of chatbots began with the creation of ELIZA in 1966 \cite{daher2020empathic}, and today they are commonly used in customer service but have limitations in showing empathy. Empathy, defined as understanding and sharing another's feelings, is deemed challenging for conversational agents. Recent research, however, indicates the feasibility of generating empathic responses in chatbots, particularly in customer service contexts, including in the healthcare, especially in providing physical health diagnosis through short text conversations. 

Several studies available in the literature to compensate for the shortcomings of lack of empathic expressions in computer and AI employed the use of chatbots to convey emptional expressions. One study used chatbots to mediate social presence and trust in consumer emotions \cite{sun2022research}, creating a sense of social presence using emoticons, appropriate language styles, and other cues that contribute to a more authentic and human-like interaction. Another study used empathic chatbot to understand users' emotional states and generate responses that convey understanding and addressing challenges in handling multi-layered, context-sensitive, and implicitly expressed emotions in text \cite{casas2021enhancing}. It utilized advanced models and tools involving a benchmark bot and an empathic bot. To enhance empathic capabilities, they fine-tuned the language model on empathic conversations, incorporating the user's emotional state into the input. The findings from the empathic bot include the effectiveness of transformer-based language models, the positive impact of training on empathic conversations, and the influence of emotional valence on perceived empathy. 

Similarly, chatbot application is finding its way into gender voice discrimination. \cite{jung2023female} explores the tension between designing empathic agents and the gender assignment of chatbots and how they can relate to the design of the metaphor of the chatbots in conversational agents (Cas). Computers lack genuine emotions, however CAs can simulate and trigger empathy by adhering to human-social rules during interactions. This simulation involves techniques like sentiment analysis, emotion detection, and mimicking emotions to enhance user engagement. Notably, the paper explores whether CAs, specifically those perceived as having feminine qualities, evoke more empathy in human-chatbot interactions compared to other gender perceptions.

\paragraph{Digitizing Human Emotions:}
As computing machines and AI need to be programmed to work, inculcating  empathy into machines is through digitizing human emotions. One study used non-invasive techniques like  electroencephalography (EEG), specifically the Emotive Epoc headset to digitize human emotions \cite{roshdy2021machine} and the primary goal was to conduct a proof-of-concept experiment enabling a humanoid robot's control through digitized emotions, emphasizing potential applications in healthcare . Their contribution lies in adapting mature image recognition tools from Artificial Neural Networks (ANNs) for emotion recognition, and the resulting Brain-Computer Interface (BCI) system is designed to enable the robot to emulate empathy and interact with subjects based on predefined behavioral models. Similarly, \cite{saffaryazdi2022emotion} explores the significance of recognizing emotions in face-to-face conversations and discussed the limitations of traditional cues like facial expressions and gestures, emphasizing the potential of analyzing brain activity and physiological signals for more reliable emotion recognition, especially in situations where emotions may be concealed. The study presents an experimental setup for inducing spontaneous emotions in conversations and creating a dataset incorporating EEG, Photoplethysmography (PPG), and Galvanic Skin Response (GSR) signals. The researchers developed an intelligent user interface for video conferencing and systems with conversational digital humans capable of recognizing and responding to emotions, using PEGCONV) dataset to train these systems to enhance their ability to understand and appropriately respond to human behavior.

Further, Robots can also respond to user emotional states. \cite{bagheri2020autonomous} enhanced human-robot interactions (HRI) by proposing an Automatic Cognitive Empathy Model (ACEM) for humanoid robots. The goal was to achieve more extended and engaging interactions by having robots respond appropriately to users' emotional states using ACEM model to continuously detects users' affective states based on facial expressions, utilizing a stacked autoencoder network trained on the RAVDESS dataset. The model generates empathic behaviors adapted to users' personalities, either in parallel or as reactive responses.

\paragraph{Empathic Systems in AR/VR:}
Empathic machine is also finding applications into wearable. In exploring the use of smart glasses equipped with an emotion recognition system to enhance human-to-human communication, with a focus on doctors and autistic adults \cite{lin2020empathics}, this study identifies the potential benefits for doctors in improving patient-doctor communication and for autistic adults facing challenges in adhering to neurotypical communication norms. The proposed system combines emotion recognition, AI, and smart glasses to provide real-time feedback on the emotional state of conversation partners. User evaluations indicate positive responses from both doctors and autistic adults, highlighting the potential for improving empathy and communication. Design considerations include customizable output preferences and addressing privacy and social impact concerns. The authors emphasize the system's potential for educational purposes and ongoing development in the context of social impact and user experience. In a related study, \cite{gupta2021towards} explores the use of VR with wearable physiological sensors to examine how recalling autobiographical memories (AM) with emotional content affects an individual's physiological state. By replicating the Autobiographical Memory Test (AMT) in VR, participants were presented with positive, negative, and neutral words to trigger memory recall. The study observed a positive influence of AM recall on electrodermal activity (EDA) peak amplitude, EDA peak number, and pupil diameter compared to situations without recall. However, emotional AM recall did not produce a significant impact. The article concludes by discussing the potential and limitations of utilizing autobiographical memory to enhance personalized mobile VR experiences in conjunction with physiological sensors. The findings reveal a significant effect of AM recall on EDA mean peak amplitude, EDA peak number, and pupil diameter compared to a no-recall condition. However, no significant differences were identified in emotional AM recall (positive, negative, neutral).

\section{Discussion}

This review identifies key advancements, limitations, and research gaps in the integration of emotional expression into speech-to-text (STT) interfaces, particularly within immersive environments such as virtual and augmented reality (VR/AR). While STT technologies have significantly improved accessibility—especially for DHH users—they typically fail to convey affective cues, resulting in reduced expressiveness and diminished communication quality~\cite{murphy1983impact}. Motivated by this persistent limitation in traditional captioning systems, our review synthesizes current literature to assess the state of research and inform future directions for more emotionally resonant captioning systems.

Through a structured narrative review of 37 peer-reviewed studies published between 2020 and 2024, we examined developments across communication enhancement technologies, captioning innovation, emotion recognition in XR, and empathic human-computer systems. Our thematic synthesis included a focused review of visual and affective augmentation strategies, which show promise for conveying emotion through ambient cues and interactive design elements~\cite{10.1145/3532525.3532527, 10.1007/978-3-031-60881-0_24, 10.1145/3411763.3451698}. However, these strategies are largely decoupled from real-time STT outputs, underscoring a significant gap in current systems: the absence of emotionally expressive, real-time captions in immersive environments.

Despite growing interest in affective computing and XR accessibility, few studies directly address the integration of emotion into STT interfaces within AR/VR contexts. Current approaches focus primarily on visual enhancements or emotion-aware avatars, without embedding affective cues directly into the transcribed text~\cite{hu2019emojilization, rashid2008}. This limitation presents a critical opportunity for innovation—particularly in education and training scenarios where emotional tone can support comprehension, reduce cognitive load, and increase learner engagement~\cite{pirker2021potential, chen2023emotion}.

Research into captioning innovations such as Rashid’s animated text framework~\cite{rashid2006expressing, rashid2008} and Hu’s emojization tool~\cite{hu2019emojilization} exemplifies early efforts to embed emotion into transcription. These methods—whether through animated captions, color cues, or emoji augmentations—have shown positive responses from DHH users~\cite{lee2007emotive}, and highlight the potential for more expressive, inclusive STT systems. Yet, scalability and vocabulary limitations remain challenges, especially for deployment in real-time XR applications.

Extending these innovations into immersive platforms offers new possibilities. Recent AR and VR captioning prototypes demonstrate technical feasibility for real-time STT with spatial awareness and keyword extraction~\cite{li2023real, alnafjan2017designing}. However, few efforts have integrated emotional nuance into these systems. In educational settings, where XR-based captioning is increasingly applied, the inclusion of affective information could enhance both accessibility and engagement~\cite{pirker2021potential, chen2023emotion}. Research on presence and affective response in VR environments further suggests that immersive platforms are well-suited for emotionally resonant experiences~\cite{kilteni2012sense}.

Beyond captioning, affective computing in adjacent domains—such as healthcare, robotics, and human-agent interaction—provides valuable insights. Studies show that emotionally responsive systems, powered by machine learning, facial recognition, and physiological sensing, can enhance empathy, personalization, and user satisfaction~\cite{wang2004communicating, cernea2015emotion, kumarapeli2022emotional}. Applications in education and clinical settings demonstrate that recognizing and responding to user emotion improves outcomes and fosters trust~\cite{schiano2000face}.
The digitization of human emotion, including efforts using brain-computer interfaces (BCIs) and affective robotics, represents a broader trend toward emotionally intelligent technology. BCI-driven systems that interpret EEG or physiological signals offer promising methods for capturing emotional states~\cite{roshdy2021machine, saffaryazdi2022emotion}, though their integration with STT remains underexplored. Similarly, empathy models like ACEM for humanoid robots illustrate how emotional awareness can be incorporated into interaction design to enable adaptive and human-like behavior~\cite{bagheri2020autonomous}.

Wearable technologies, such as smart glasses with built-in emotion recognition, have also shown promise in improving communication among autistic individuals and between doctors and patients~\cite{lin2020empathics}. These tools deliver real-time emotional feedback to enhance interpersonal understanding, and their integration into educational contexts could support more personalized and empathetic learning experiences.
Multisensory XR systems that combine physiological data, autobiographical memory, and real-time feedback offer new frontiers for emotionally responsive interaction. Studies exploring such experiences emphasize the potential for tailoring virtual environments to individual affective states~\cite{gupta2021towards}, providing valuable frameworks for the design of emotionally expressive captioning systems.

Overall, the findings of this review suggest a compelling case for interdisciplinary research that bridges affective computing, psychophysiology, XR design, and accessibility. The future of inclusive STT systems lies in integrating emotional expression not as a secondary visual layer, but as an intrinsic component of the transcription process—transforming captions from neutral information displays into emotionally meaningful communication tools.

In sum, advancing emotional expression in STT systems holds particular promise for enhancing communication accessibility for DHH users. By embedding emotion-aware cues into real-time captions, especially within XR environments, designers and researchers can move toward more human-centered, empathetic technologies. This work underscores the importance of inclusivity not only in functionality but in emotional fidelity—ensuring that all users, regardless of hearing ability, can access the full depth of spoken communication.

\subsection{Limitations}
The limitations of this study were: (i) it was a rapid review whose purpose is to survey existing work in emotional expressions, find existing gaps and inspire further research. (ii) the narrative review did not follow a strict rule expected of systematic literature review, hence could have missed some relevant papers. However, the papers that were found eligible and used in this review are the most relevant. (iii) the review is limited to peer-reviewed papers that were in English.

\subsection{Gaps, Challenges and Opportunities}

\begin{enumerate}
\item Innovations in emotive captioning and speech emotion recognition warrant further exploration, particularly in educational settings where DHH students and other learners rely on STT as their main method of communication during lectures.

\item Further integration of cognitive empathy models is needed to improve emotion detection in short conversations. Enhancements in medical chatbots show potential for empathy-driven systems \cite{daher2020empathic}.

\item There is a need for automated metrics to assess empathy in text-based systems. Future studies should also refine definitions of empathy and explore emotion intensity and categories, while addressing ethical concerns in human-like chatbot design \cite{casas2021enhancing}.

\item Research should explore alternative empathic response markers, including diverse character styles and physiological indicators, and test these in immersive environments like VR \cite{higgins2023investigating}.

\item Emotion recognition systems should incorporate multimodal inputs, including EEG and physiological signals, for a holistic approach to understanding affective states in conversation \cite{saffaryazdi2022emotion}.

\item Future investigations should assess applications of emotion-aware systems in educational contexts, including medical education, and evaluate their integration into training curricula \cite{lin2020empathics}.

\item Larger-scale studies are needed with more emotionally intense stimuli and diverse biosignals, applicable to VR learning, exposure therapy, and adaptive gaming \cite{gupta2021towards}.

\item The field should prioritize real-time integration of emotional visualization techniques with ASR captions in XR, moving beyond ambient indicators to emotion-aware captions that directly reflect STT output.
\end{enumerate}

\subsection{Future Work Directions}

To address the outlined challenges and research gaps, we propose the following directions for future exploration:

\begin{enumerate}
\item \textbf{Integrating Visual Augmentation in AR/VR STT Systems:} Investigate how the use of emojis, animated avatars, and symbolic visuals can enhance emotional expression in real-time captions.

\item \textbf{Designing Emotionally Resonant Caption Formats:} Identify optimal caption formats and placement strategies that preserve speaker affect and reduce viewer distraction in immersive settings.

\item \textbf{Reducing Cognitive Load through Personalization:} Develop adaptive captioning interfaces that adjust visual emphasis, pacing, and layout based on user preferences and cognitive workload.

\item \textbf{Dynamic Substitution of Emotive Language:} Explore mechanisms for substituting emotionally expressive language with visual or symbolic equivalents (e.g., icons, colors) during live transcription.

\item \textbf{Cross-Platform Evaluation Frameworks:} Build comprehensive evaluation frameworks that assess the emotional clarity, usability, and accessibility of captioning strategies across XR platforms.
\end{enumerate}

\section{Conclusion}

This narrative review explored the integration of emotional expression into STT interfaces within immersive XR environments, emphasizing the intersection of psychophysiology, accessibility, and user-centered design. While STT technologies have significantly advanced accessibility for DHH users, the omission of emotional sues in transcribed speech remains a critical barrier to natural and expressive communication.

Our findings highlight a growing recognition in HCI of the need for emotion-aware captioning, with emerging strategies such as emotion-driven avatars, color-coded overlays, and ambient cues offering promising—but largely disconnected—approaches. A key research gap lies in the lack of direct integration between real-time STT outputs and affective visualization techniques.

Addressing this gap requires interdisciplinary collaboration across affective computing, neuroscience, and XR systems design. Future research should investigate the effectiveness of real-time emotional augmentation—such as emojis, avatars, and dynamic text formatting—in reducing cognitive load and enhancing comprehension in AR/VR settings. This review highlights the urgent need to move beyond neutral captioning toward systems that can capture and convey the emotional context of speech.

By bridging accessibility with emotional expressiveness, the next generation of STT systems can support more inclusive, empathetic, and engaging communication—paving the way for richer user experiences in immersive environments and beyond.

\end{document}